\documentclass[twocolumn]{jpsj3} 

\usepackage{color}

\title{
Effects of Lattice and Molecular Phonons on Photoinduced Neutral-to-Ionic Transition Dynamics in Tetrathiafulvalene-$p$-Chloranil 
}

\author{
Kenji \textsc{Yonemitsu}$^{1,2,3}$\thanks{E-mail: kxy@ims.ac.jp}
}

\inst{
$^{1}$Institute for Molecular Science, Okazaki, Aichi 444-8585, Japan \\
$^{2}$Department of Functional Molecular Science, Graduate University for Advanced Studies, Okazaki, Aichi 444-8585, Japan \\
$^{3}$JST, CREST, Tokyo 102-0075, Japan
}

\abst{
For electronic states and photoinduced charge dynamics near the neutral-ionic transition in the mixed-stack charge-transfer complex tetrathiafulvalene-$p$-chloranil (TTF-CA), we review the effects of Peierls coupling to lattice phonons modulating transfer integrals and Holstein couplings to molecular vibrations modulating site energies. The former stabilizes the ionic phase and reduces discontinuities in the phase transition, while the latter stabilizes the neutral phase and enhances the discontinuities. To reproduce the experimentally observed ionicity, optical conductivity and photoinduced charge dynamics, both couplings are quantitatively important. In particular, strong Holstein couplings to form the highly-stabilized neutral phase are necessary for the ionic phase to be a Mott insulator with large ionicity. A comparison with the observed photoinduced charge dynamics indicates the presence of strings of lattice dimerization in the neutral phase above the transition temperature. 
}

\kword{
photoinduced phase transition, neutral-ionic transition, electron-lattice interaction, electron-molecular-vibration coupling
}

\begin{document}
\maketitle

\section{Introduction}

Nonequilibrium electronic states have drawn attention as a field where novel electronic functions and properties are being searched for. Correlated electron and electron-phonon systems with rich electronic phase diagrams are promising because synergy can be exploited, which is inherent to respective orders at low temperatures. Among nonequilibrium phenomena, photoinduced phase transitions occur synergistically since a low density of photons relative to the density of molecules or atoms can alter the electronic phase realized in such a system. A variety of photoinduced phase transitions are known to proceed very rapidly. \cite{tokura_jpsj06,yonemitsu_pr08} 

From the viewpoint of controlling functions in nonequilibrium states, it is important to learn a means by which a photoinduced phase transition proceeds. The most fundamental information is on relevant interactions that stabilize the ground state. It is generally true that, even if a few interaction parameters are sufficient to describe equilibrium properties, they are still insufficient for describing nonequilibrium dynamics. Additional interactions can be crucial in determining whether and how the phase transition is photoinduced. Among organic materials that have been regarded as strongly correlated electron systems, some have been realized to have substantial electron-molecular-vibration (EMV) couplings that contribute to the stabilization of the ground state and photoinduced dynamics. 

For instance, the charge order in the quasi-two-dimensional organic salt $ \alpha $-(BEDT-TTF)$_2$I$_3$ [BEDT-TTF=bis(ethylenedithio)tetrathiafulvalene] is basically driven by a long-range Coulomb interaction. \cite{mckenzie_s97,seo_jpsj00} Relatively weak Peierls couplings to lattice phonons or molecular rotations are important for understanding the interrelation between the altered charge distribution and the structural deformation \cite{tanaka_jpsj08,miyashita_jpsj08} as well as the long-time behavior of photoinduced dynamics. \cite{tanaka_jpsj10,miyashita_jpsj10} From the early-stage dynamics after photoexcitation, however, substantially strong Holstein couplings to molecular vibrations are evident in the Fano destructive interference between vibrations containing C=C stretching and correlated electrons' motion. \cite{kawakami_prl10} These molecular vibrations stabilize the charge order to a substantial extent. 

Most organic compounds that show electronic phase transitions have C=C bonds in their constituent molecules. Their vibration frequencies are well known to depend on the charged state of the molecule. It is therefore expected that EMV couplings are generally strong even though the electronic state has strong electron correlations in terms of intermolecular spin and charge correlation functions. In this study, we focus on the mixed-stack charge-transfer complex TTF-CA, whose thermal and photoinduced neutral-ionic transitions have intensively been studied both experimentally \cite{torrance_prl81a,torrance_prl81,koshihara_prb90,lecointe_prb95,koshihara_jpcb99,suzuki_prb99,iwai_prl02,collet_s03,guerin_cp04,tanimura_prb04,okamoto_prb04,matsuzaki_jpsj05,davino_prl07,guerin_prl10,uemura_prl10} and theoretically. \cite{nagaosa_jpsj86b,nagaosa_jpsj86c,painelli_prb88,toyozawa_ssc92,sakano_jpsj96,huai_jpsj00,pati_prb01,kawamoto_prb01,yonemitsu_prb02,delfreo_prl02,miyashita_jpsj03,kishine_prb04,yonemitsu_prb06,iwano_prl06,soos_prb07} We review the origin of the discontinuities in ionicity and optical conductivity during the neutral-ionic transition. Their quantitative understanding will become important for controlling photoinduced phase transition dynamics. 

\section{Discontinuities during Neutral-Ionic Transition}

The importance of the intersite Coulomb repulsion $ V $ for discontinuities during the neutral-ionic transition has been pointed out in ref.~\citen{nagaosa_jpsj86b}. Without coupling to lattice phonons, i.e., without dimerization, the spin gap at the phase boundary is zero on the ionic (i.e., regular-Mott-insulator) side and $ V $ on the neutral (i.e., band-insulator) side in the limit of vanishing transfer integrals. Here, the creation of a neutral-ionic domain wall requires $ V/2 $ of energy, which allows for the phase separation of neutral and ionic domains at the boundary. The transfer integral $ t_0 $ introduces quantum fluctuations. As long as $ 2 t_0 $ is smaller than $ V $, however, the creation energy remains finite. The discontinuous neutral-ionic transition is demonstrated by quantum Monte Carlo simulations. \cite{nagaosa_jpsj86b}

The ionic phase is known to have dimerization. The strongest candidate for its mechanism is due to the spin-Peierls instability. \cite{nagaosa_jpsj86c} A regular ionic phase is a one-dimensional half-filled paramagnetic Mott insulator, so that it can be described by a spin-1/2 antiferromagnetic Heisenberg chain. If the superexchange interaction is modulated by lattice phonons, the system is stabilized by dimerization that produces a finite spin gap. The dimerization in the ionic phase is demonstrated by quantum Monte Carlo simulations. \cite{nagaosa_jpsj86c} As lattice phonons are strongly dimerized, i.e., as the electron-lattice coupling is strengthened, the phase boundary is shifted because only the ionic phase is stabilized, which decreases the ionicity on the ionic side of the phase boundary. The discontinuity in ionicity is strongly reduced by the electron-lattice coupling. As a consequence, the character of the ionic phase as a Mott insulator is weakened, and the transition becomes close to a continuous Peierls transition. 

For the discontinuous ionicity change, Painelli and Girlando have pointed out that both EMV couplings and intersite Coulomb repulsion $ V $ are important and that they collaborate. \cite{painelli_prb88} For finite systems, they perform exact diagonalization in the limit of a large on-site repulsion $ U $ and the second-order perturbation theory with respect to the electron-lattice (i.e., Peierls) and EMV (i.e., Holstein) couplings. The electron number configuration is roughly described as 2020 in the neutral phase and as 1111 in the ionic phase, so that the Holstein couplings stabilize the neutral phase, which are in contrast to the Peierls coupling that stabilizes the ionic phase through dimerization. 

As noted above, both of the Peierls and Holstein couplings are suggested to be necessary for the reproduction of ionicity and optical conductivity near the neutral-ionic phase boundary. In spite of this, the Holstein couplings have often been ignored for the following reason. Without the Peierls coupling, the bond-charge densities cannot be dimerized, and the ionic phase cannot have the ferroelectric order. In contrast, the Holstein couplings only enlarge the site energy difference between the neighboring orbitals in the neutral phase, so that they can be absorbed into the renormalized site energy difference in the adiabatic limit. The situation is quite similar to a phase transition described within the mean field theory in the sense that the Holstein couplings stabilize the neutral phase in a synergistic manner and enhance the discontinuity during the transition. Namely, the degrees of renormalization are different between the two phases. 

In this paper, we include both couplings in the model. The charge dynamics after photoexcitation of the neutral phase shows a large contribution from molecular vibrations. This is consistent with the ultrafast charge and molecular-vibration dynamics recently observed by the transient reflectivity measurements. \cite{uemura_prl10} Numerical calculations show that the initial neutral state must be accompanied with a small dimerization to reproduce the experimentally observed, coherently oscillating dimerization immediately after photoexcitation. The presence of strings of lattice dimerization in the neutral phase has indeed been observed by the latest X-ray diffuse scattering. \cite{guerin_prl10} 

\section{Model with Lattice and Molecular Phonons}

We use the one-dimensional half-filled extended ionic Hubbard-Peierls-Holstein model 
\begin{eqnarray}
H & = & 
-\sum_{j\sigma}[ t_0 - \alpha (u_{j+1}-u_j) ]
( c_{j\sigma}^\dagger c_{j+1\sigma} + c_{j+1\sigma}^\dagger c_{j\sigma} ) 
\nonumber \\ & & 
+\frac{\Delta}{2} \sum_{j\sigma} (-1)^j c_{j\sigma}^\dagger c_{j\sigma} 
-\sum_{mj\sigma} \beta_j^{(m)} v_j^{(m)} c_{j\sigma}^\dagger c_{j\sigma} 
\nonumber \\ & & 
+U \sum_j n_{j\uparrow} n_{j\downarrow} +V \sum_j n_j n_{j+1} 
\nonumber \\ & & 
+\sum_j \left[ \frac{K_\alpha}{2} ( u_{j+1}-u_j )^2 
+ \frac{2K_\alpha}{\omega_\alpha^2} \dot{u}_j^2 \right] 
\nonumber \\ & & 
+\sum_{mj} \left[ \frac{K_{\beta j}^{(m)}}{2} v_j^{(m)2} 
+ \frac{K_{\beta j}^{(m)}}{2\omega_{\beta j}^{(m)2}} \dot{v}_j^{(m)2} \right] 
\;, \label{eq:model}
\end{eqnarray}
where $ c^\dagger_{j\sigma} $ creates an electron with spin $ \sigma $ at site $ j $, $ n_{j\sigma} $=$ c_{j\sigma}^\dagger c_{j\sigma} $, and $ n_j $=$ \sum_\sigma n_{j\sigma} $. The parameter $ t_0 $ denotes the transfer integral on a regular lattice (i.e., without lattice distortion), $ \Delta $ the site energy difference between neighboring orbitals when molecular distortions are absent, $ U $ the on-site repulsion strength, and $ V $ the nearest-neighbor repulsion strength. The lattice displacement $ u_j $ at site $ j $ modulates the transfer integral between the $ (j-1) $th and $ j $th orbitals and that between the $ j $th and $ (j+1) $th orbitals with the coefficient $ \mp \alpha $. The displacement $ v_j^{(m)} $ in the $ m $th mode on the $ j $th molecule modulates the site energy with the coefficient $ \beta_j^{(m)} $. The quantities $ \dot{u}_j $ and $ \dot{v}_j^{(m)} $ are the time derivatives of $ u_j $ and $ v_j^{(m)} $, respectively. The parameters $ K_\alpha $ and $ K_{\beta j}^{(m)} $ are their elastic coefficients, and $ \omega_\alpha $ and $ \omega_{\beta j}^{(m)} $ are their bare phonon energies, respectively. 

For these model parameters, we take eV as the unit of energy and use $ t_0 $=0.17, \cite{yonemitsu_prb06} $ U $=1.5, and $ V $=0.6; we vary $ \Delta $ around the phase boundary depending on the Peierls and Holstein coupling strengths. We define the strengths of these couplings as $ \lambda_\alpha \equiv \alpha^2/K_\alpha  $ and $ \lambda_{\beta j} \equiv \sum_m \beta_j^{(m)2}/K_{\beta j}^{(m)} $. As long as the displacements are treated classically, the ground state is given by their static configuration, and the ground state is determined not by the distribution of $ \beta_j^{(m)2}/K_{\beta j}^{(m)} $, but by their sum, $ \lambda_{\beta j} $. \cite{yonemitsu_prb09} The displacements are scaled using $ \alpha $=$ \beta_j^{(m)} $=1, so that we have $ \lambda_\alpha = 1/K_\alpha  $ and $ \lambda_{\beta j} = \sum_m 1/K_{\beta j}^{(m)} $. For simplicity, we set $ \lambda_{\beta 2i-1} = \lambda_{\beta 2i} = \lambda_\beta $. 

As for phonons, we take one mode for the donor molecule and two modes for the acceptor molecule in addition to the lattice phonon mode, and use parameters that approximately reproduce the experimentally observed phonon energies: $ \omega_\alpha $=0.013 (this energy is softened to be about 0.0066 when strongly coupled to electrons), $ \omega_{\beta 2i}^{(1)} \equiv  \omega_{\beta \mathrm{A}1} $=0.040, $ \omega_{\beta 2i-1}^{(1)} \equiv  \omega_{\beta \mathrm{D}} $=0.055, and $ \omega_{\beta 2i}^{(2)} \equiv  \omega_{\beta \mathrm{A}2} $=0.12. Donor and acceptor molecules are specified by odd and even $ j $'s, respectively. For simplicity, we set $ K_{\beta 2i}^{(1)} $=$ K_{\beta 2i}^{(2)} $ because the conclusion does not depend on the detailed distribution of EMV-coupling strengths. The displacements  $ v_\mathrm{A1} $, $ v_\mathrm{D} $, and $ v_\mathrm{A2} $ correspond to CA's $ a_g \nu_5 $ (320 cm$^{-1}$), TTF's $ a_g \nu_6 $ (438 cm$^{-1}$), and CA's $ a_g \nu_3 $ (957 cm$^{-1}$) modes, respectively, in ref.~\citen{uemura_prl10}, whereas we ignore TTF's $ a_g \nu_5 $ (740 cm$^{-1}$) mode because its coupling to electrons is very weak. 

Photoexcitation is introduced through the Peierls phase 
\begin{equation}
c^\dagger_{i\sigma} c_{j\sigma}
\rightarrow
e^{(ie/\hbar c) (j-i) A(t)}
c^\dagger_{i\sigma} c_{j\sigma}
\;. \label{eq:Peierls}
\end{equation}
The time-dependent vector potential $ A(t) $ for a pulse of an oscillating electric field is given by 
\begin{equation}
A(t)=
\frac{ F }{\omega_\mathrm{pmp}}
\cos(\omega_\mathrm{pmp} t)
\frac1{\sqrt{2\pi}T_\mathrm{pmp}}
\exp \left( -\frac{t^2}{2 T_\mathrm{pmp}^2} \right)
\;, \label{eq:Vector_potential}
\end{equation}
where $ \omega_\mathrm{pmp} $ is the excitation energy, $ T_\mathrm{pmp} $ is the pulse width, and $ F $ is the electric field amplitude. 

The time-dependent Schr\"odinger equation for the exact many-electron wave function on the chain of $ N $=12 sites with periodic boundary condition is numerically solved by expanding the exponential evolution operator with a time slice $ dt $=0.02 eV$^{-1}$ to the 15th order and by checking the conservation of the norm. \cite{yonemitsu_prb09} The initial state is set in the electronic ground state. The classical equations for the lattice and molecular displacements are solved by the leapfrog method, where the forces are derived from the Hellmann-Feynman theorem: 
\begin{eqnarray}
& & 
\frac{4K_\alpha}{\omega_\alpha^2} 
\frac{\mathrm{d}^2 u_j}{\mathrm{d}t^2} = 
K_\alpha ( u_{j+1} -2 u_j + u_{j-1}) 
\nonumber \\ & & \ \ 
+ \alpha \sum_\sigma \langle 
c_{j\sigma}^\dagger c_{j+1\sigma} + \mathrm{H.c.} 
- c_{j-1\sigma}^\dagger c_{j\sigma} - \mathrm{H.c.} 
\rangle
\;, \label{eq:Newton_alpha}
\end{eqnarray}
\begin{equation}
\frac{K_{\beta j}^{(m)}}{\omega_{\beta j}^{(m)2}} 
\frac{\mathrm{d}^2 v_j^{(m)}}{\mathrm{d}t^2} = 
- K_{\beta j}^{(m)} v_j^{(m)} + 
\beta_j^{(m)} \sum_\sigma \langle c_{j\sigma}^\dagger c_{j\sigma} \rangle 
\;. \label{eq:Newton_beta}
\end{equation}
Unless otherwise stated, the initial displacements are at the minimum adiabatic potential for the electronic ground state, so that the initial velocities are zero. 

\section{Ground-State Properties}

\begin{figure}
\includegraphics[height=9.6cm]{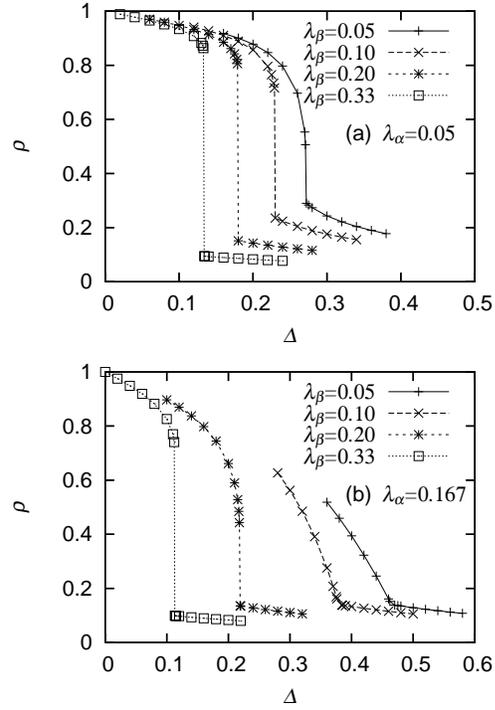}
\caption{
Ionicity $ \rho $ as a function of site energy difference $ \Delta $ for (a) weak Peierls coupling $ \lambda_\alpha $=0.05, and (b) strong Peierls coupling $ \lambda_\alpha $=0.167 with different strengths of $ \lambda_\beta $. 
\label{fig:rho_delta_alpha}}
\end{figure}

\begin{figure}
\includegraphics[height=9.6cm]{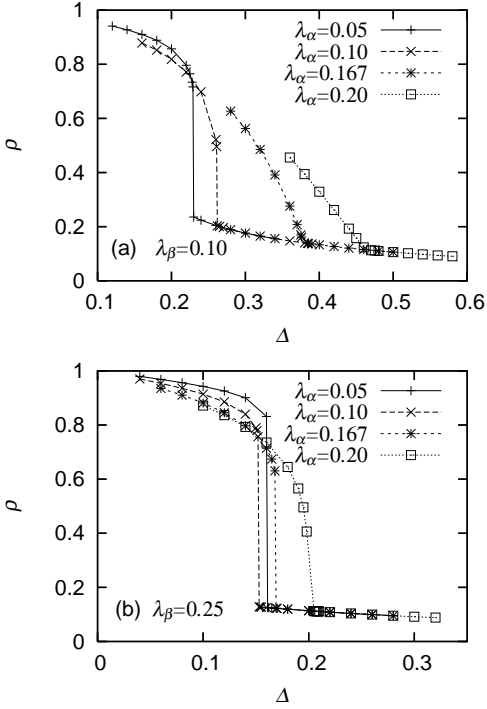}
\caption{
Ionicity $ \rho $ as a function of site energy difference $ \Delta $ for (a) weak Holstein coupling $ \lambda_\beta $=0.10, and (b) strong Holstein coupling $ \lambda_\beta $=0.25 with different strengths of $ \lambda_\alpha $. 
\label{fig:rho_delta_beta}}
\end{figure}

Figures~\ref{fig:rho_delta_alpha} and \ref{fig:rho_delta_beta} show the ionicity 
\begin{equation}
\rho \equiv 1 + (1/N) \sum_{j=1}^N (-1)^j \langle n_j \rangle 
\label{eq:ionicity}
\end{equation}
as a function of the site energy difference $ \Delta $ near the phase boundary with different combinations of $ \lambda_\alpha $ and $ \lambda_\beta $. 
In the vicinity of the phase boundary, one phase is stable and the other is metastable. By comparing their energies, only the ionicity in the stable phase is plotted. It is clearly shown that, as $ \lambda_\beta $ increases, the neutral phase is stabilized, the discontinuity in ionicity is enlarged, the ionicity in the neutral phase on the large-$ \Delta $ side of the phase boundary becomes smaller, and the ionicity in the ionic phase on the small-$ \Delta $ side becomes larger (Fig.~\ref{fig:rho_delta_alpha}). As a consequence, in order for the ionic phase to be a typical Mott insulator, i.e., in order for the ionic phase to have nearly one electron per site, $ \lambda_\beta $ should be so large that the neutral phase is sufficiently stabilized. In particular, when $ \lambda_\alpha $ is large, $ \lambda_\beta $ needs to be large [Fig.~\ref{fig:rho_delta_alpha}(b)]. Otherwise, the ionicity is too small even in the ionic phase. 

Figure~\ref{fig:rho_delta_beta} has essentially the same information as Fig.~\ref{fig:rho_delta_alpha}. It is evident that, as $ \lambda_\alpha $ increases, the ionic phase is stabilized, and the discontinuity at the transition is suppressed. When $ \lambda_\beta $ is large, the phase boundary does not shift monotonically as a function of $ \lambda_\alpha $ [Fig.~\ref{fig:rho_delta_beta}(b) and $ \lambda_\beta $=0.33 (not shown)]. This is caused by the competition between the bond charge $ \sum_\sigma \langle c_{j\sigma}^\dagger c_{j+1\sigma} + c_{j+1\sigma}^\dagger c_{j\sigma} \rangle $ localized by $ \lambda_\alpha $ and the site charge $ \sum_\sigma \langle c_{j\sigma}^\dagger c_{j\sigma} \rangle $ localized by $ \lambda_\beta $. The coupling $ \lambda_\alpha $ stabilizes only the ionic phase, while $ \lambda_\beta $ stabilizes both phases (i.e., the neutral phase strongly and the ionic phase weakly). When $ \lambda_\alpha $ is small and the dimerization is weak (i.e., when lattice phonons only slightly lower the energy of the ionic phase), the increasing $ \lambda_\alpha $ gains energy by localizing the bond charge, but it loses more energy by delocalizing the site charge. As a consequence, for large $ \lambda_\beta $, the intermediate $ \lambda_\alpha $ stabilizes the neutral phase more strongly than the ionic phase. This counterintuitive result cannot be obtained using the second-order perturbation theory. \cite{painelli_prb88} Nevertheless, the shift of the phase boundary is small in this case. 

\begin{figure}
\includegraphics[height=14.4cm]{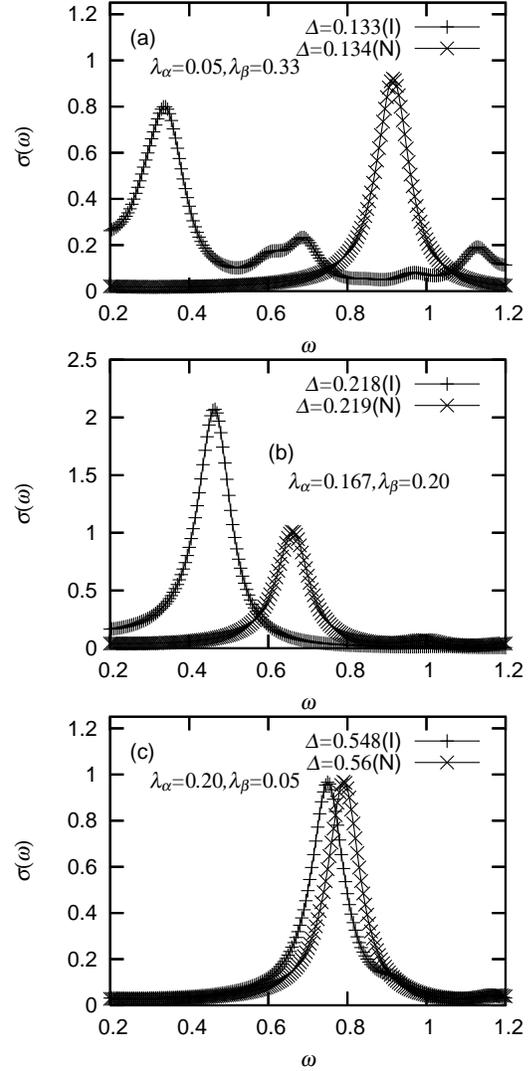}
\caption{
Optical conductivity spectra in ionic (with smaller $ \Delta $) and neutral (with larger $ \Delta $) phases near the phase boundary with (a) $ \lambda_\alpha $=0.05 and $ \lambda_\beta $=0.33, (b) $ \lambda_\alpha $=0.167 and $ \lambda_\beta $=0.20, and (c) $ \lambda_\alpha $=0.20 and $ \lambda_\beta $=0.05. 
\label{fig:conductivity_alpha_beta}}
\end{figure}

Experimentally, it is known that the thermal transition from the neutral to ionic phases is accompanied by a large ionicity jump from 0.3 to 0.7. \cite{koshihara_prb90} This indicates a large $ \lambda_\beta $. Of course, a finite $ \lambda_\alpha $ is certain to induce dimerization and the three-dimensional ferroelectric order with a broken inversion symmetry. \cite{lecointe_prb95} As shown below, a large $ \lambda_\alpha $ is also necessary for the reproduction of the optical conductivity alteration during the transition [Fig.~\ref{fig:conductivity_alpha_beta}(b)]. Therefore, both $ \lambda_\alpha $ and $ \lambda_\beta $ are large in TTF-CA. 

The optical conductivity $ \sigma(\omega) $ in the ground state $ \mid \! \psi_0 \rangle $ is calculated using 
\begin{equation}
\sigma(\omega) = -\frac{1}{N\omega} \mathrm{Im} 
\langle \psi_0 \! \mid \! j 
\frac{1}{\omega + i \epsilon + E_0 - H}
j \! \mid \! \psi_0 \rangle
\;, \label{eq:conductivity}
\end{equation}
where $ j $ is the current operator $ j \equiv - \partial H/\partial A $, $ \epsilon $ is a peak-broadening parameter set at 0.05, and $ E_0 = \langle \psi_0 \! \mid \! H \! \mid \! \psi_0 \rangle $. Figure~\ref{fig:conductivity_alpha_beta} shows $ \sigma(\omega) $ in both phases close to the phase boundary. 
When only $ \lambda_\beta $ is large, $ \sigma(\omega) $ is largely altered during the transition [Fig.~\ref{fig:conductivity_alpha_beta}(a)]. The optical gap in the neutral phase strongly stabilized by $ \lambda_\beta $ is large, while that in the ionic phase is small in the present choice of $ U $ for TTF-CA. Experimentally, such a large alteration is not observed at the transition. When only $ \lambda_\alpha $ is large, the discontinuities in physical quantities at the transition are suppressed: the $ \sigma(\omega) $ spectra in both phases are similar [Fig.~\ref{fig:conductivity_alpha_beta}(c)]. 

\section{Photoinduced Dynamics from Ground State}

\begin{figure}
\includegraphics[height=19.2cm]{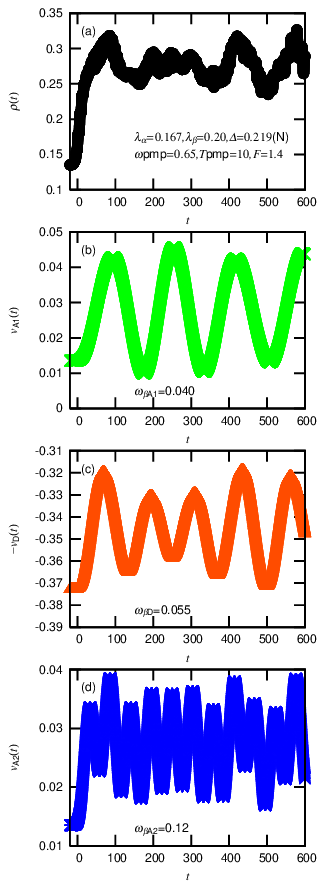}
\caption{(Color online)
Transient quantities during and after charge-transfer photoexcitation of neutral phase using $ \omega_\mathrm{pmp} $=0.65, $ T_\mathrm{pmp} $=10, and $ F $=1.4 in case of strong Holstein and Peierls couplings $ \lambda_\alpha $=0.167 and $ \lambda_\beta $=0.20: 
(a) ionicity $ \rho(t) $ and displacements 
(b) $ v_\mathrm{A1}(t) $ with bare energy $ \omega_{\beta\mathrm{A1}} $=0.040 for the acceptor molecule, 
(c) $ -v_\mathrm{D}(t) $ with bare energy $ \omega_{\beta\mathrm{D}} $=0.055 for the donor molecule, and 
(d) $ v_\mathrm{A2}(t) $ with bare energy $ \omega_{\beta\mathrm{A2}} $=0.12 for the acceptor molecule. 
\label{fig:evolution}}
\end{figure}

In the case of a large $ \lambda_\alpha $ and a large $ \lambda_\beta $ [Fig.~\ref{fig:conductivity_alpha_beta}(b)], the neutral phase near the phase boundary is photoexcited with an energy $ \omega_\mathrm{pmp} $=0.65 just above the optical gap. The time evolution of the ionicity $ \rho(t) $ during and after photoexcitation is plotted in Fig.~\ref{fig:evolution}(a). For comparison, the displacement on the acceptor molecule $ v_\mathrm{A1}(t) $ (with a lower $ \omega_{\beta\mathrm{A}} $) is shown in Fig.~\ref{fig:evolution}(b), the displacement on the donor molecule $ -v_\mathrm{D}(t) $ [$(-1)$ is multiplied so as to oscillate in the same phase with $ \rho(t) $.] in Fig.~\ref{fig:evolution}(c), and the displacement on the acceptor molecule $ v_\mathrm{A2}(t) $ (with a higher $ \omega_{\beta\mathrm{A}} $) in Fig.~\ref{fig:evolution}(d). 
As $ \rho(t) $ increases, the electron density increases for the acceptor molecule and decreases for the donor molecule [eq.~(\ref{eq:ionicity})], so that the displacement increases at the acceptor molecule and decreases at the donor molecule [eq.~(\ref{eq:Newton_beta})]. Thus, the quantities shown in Figs.~\ref{fig:evolution}(b)-\ref{fig:evolution}(d) basically behave as $(-1)$ times the cosine function. The ionicity $ \rho(t) $ receives a positive feedback from these molecular displacements and oscillates in the same phase with them. This $-$cosine behavior is consistent with the experimental observation, \cite{uemura_prl10} demonstrating that the neutral phase is stabilized by EMV couplings. 

To investigate quantum effects, we treated the molecular vibration with the highest phonon energy quantum-mechanically as in ref.~\citen{kawakami_prl10}. The difference between the wave profile when classically treated and that when quantum-mechanically treated was small (not shown). This is in contrast to the case reported in ref.~\citen{kawakami_prl10}, where the photoinduced transition is from an insulator to a metal, so that low-energy electronic excitations exist and interfere quantum-mechanically with phonons. In such a case, the difference is large. In the present case, where the photoinduced transition is from an insulator to another insulator, the electronic excitations maintain a gap much larger than the phonon energies. The wavelet analysis does not show a trace of quantum interference. 

In the present calculation, we did not adjust the distribution of $ \beta_j^{(m)2}/K_{\beta j}^{(m)} $ so as to fit to the experimental data. Therefore, we can only roughly compare the experimental \cite{uemura_prl10} and theoretical results for contribution of molecular vibrations to the photoinduced ionicity modulation. They are comparable in that the oscillation amplitude in the photoinduced ionicity change amounts to approximately one third in the present calculation and one fifth in the experimental result. We have calculated the photoinduced dynamics for different combinations of ($ \lambda_\alpha $, $ \lambda_\beta $), (0.167, 0.2), (0.167, 0.25), (0.167, 0.33), and (0.2, 0.33), all of which roughly reproduce the discontinuities in ionicity and optical conductivity at the transition. In these cases, the above contribution of molecular vibrations is roughly one third. 

However, there is a qualitative difference between the experimentally observed and present numerical results, which originates from the fact that the initial state here is the ground state with the regular and static lattice configuration. It does not contain thermal fluctuations. In the experiment, the lattice phonons are ready to be dimerized. \cite{uemura_prl10} The photoinduced ionicity dynamics has a large contribution from their slow oscillation, which fits to a damped oscillator. In the present case, the inversion symmetry of the neutral phase is broken by the oscillating electric field, so that lattice phonons begin to dimerize. However, its growth rate is quite low because it essentially corresponds to the spontaneous broken symmetry. The dimerization becomes significant only after t=600 in Fig.~\ref{fig:evolution}. If the initial state contains random numbers in $ u_j $, $ v_j^{(m)} $, $ \dot{u}_j $, and $ \dot{v}_j^{(m)} $, according to the Boltzmann distribution at a finite temperature of 0.01 eV, these fluctuations accelerate the growth of $ (-1)^j u_j $, but they do not oscillate it like $(-1)$ times the cosine function as experimentally observed. It should be noted that this calculation cannot simulate thermally induced domains larger than the present system. 

\section{Dynamics with Precursory Dimerization}

\begin{figure}
\includegraphics[height=14.4cm]{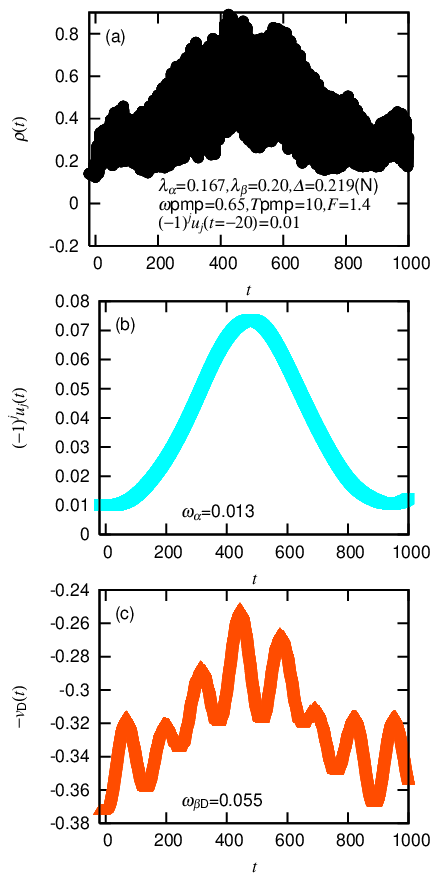}
\caption{(Color online)
Transient quantities after setting initial dimerization $ (-1)^j u_j (t=-20)=0.01 $ and charge-transfer photoexcitation of neutral phase using $ \omega_\mathrm{pmp} $=0.65, $ T_\mathrm{pmp} $=10, and $ F $=1.4 in same case as that in Fig.~\ref{fig:evolution}: 
(a) ionicity $ \rho(t) $, 
(b) dimerization $ (-1)^j u_j (t) $, and  
(c) displacement $ -v_\mathrm{D}(t) $ with bare energy $ \omega_{\beta\mathrm{D}} $=0.055 for the donor molecule. 
\label{fig:shift_evolution}}
\end{figure}

In this section, we introduce in the initial state a dimerization $ (-1)^j u_j $=0.01, which is much smaller than $ (-1)^j u_j $=0.058 of the ground state at $ \Delta $=0.218 on the ionic side of the phase boundary. After obtaining the electronic ground state with fixed $ (-1)^j u_j $=0.01, we take it as the initial state and apply to it a pulse of an oscillating electric field with energy $ \omega_\mathrm{pmp} $=0.65 again. The time evolution of the ionicity $ \rho(t) $ is plotted in Fig.~\ref{fig:shift_evolution}(a). For comparison, the dimerization $ (-1)^j u_j(t) $ is shown in Fig.~\ref{fig:shift_evolution}(b), and the displacement on the donor molecule $ -v_\mathrm{D}(t) $ in Fig.~\ref{fig:shift_evolution}(c). 
Now, $ (-1)^j u_j $ rapidly increases with $ \rho(t) $ and oscillates like $(-1)$ times the cosine function. Thus, $ \rho(t) $ receives a positive feedback from $ (-1)^j u_j $ from an early stage. It is also clear that the oscillations of $ v_j^{(m)} $ are temporally modulated by $ (-1)^j u_j $, as observed experimentally. \cite{uemura_prl10} In our calculation, the molecular vibrations and lattice phonons are indirectly coupled through electrons. 

\begin{figure}
\includegraphics[height=4.8cm]{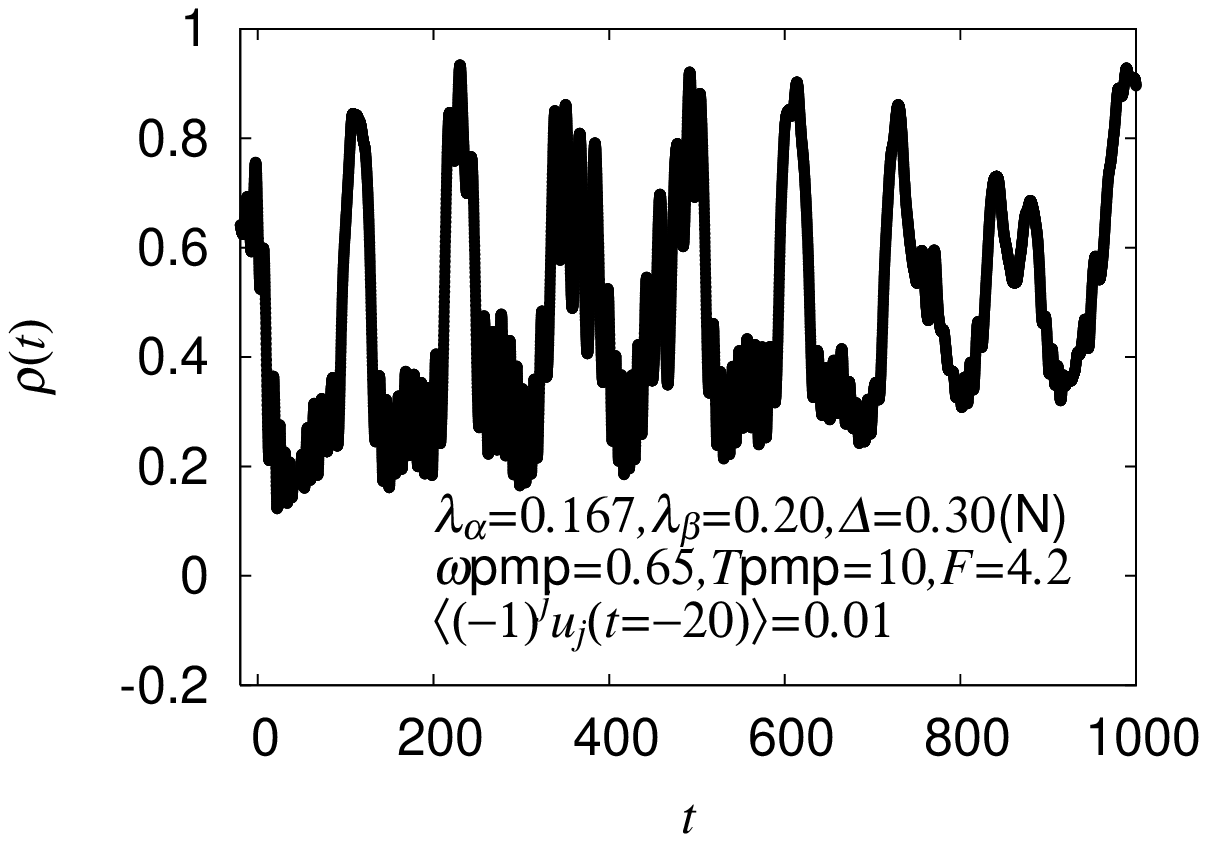}
\caption{
Transient ionicity $ \rho(t) $ after setting initial dimerization $ (-1)^j u_j (t=-20)=0.01 $, adding random numbers to phonon variables as explained in text, and charge-transfer photoexcitation of neutral phase using $ \omega_\mathrm{pmp} $=0.65, $ T_\mathrm{pmp} $=10, and $ F $=4.2 in same case as that in Fig.~\ref{fig:evolution}, but for $ \Delta $=0.30. 
\label{fig:random_shift_evol}}
\end{figure}

In this calculation, fluctuations are not introduced into $ u_j $, $ v_j^{(m)} $, $ \dot{u}_j $, or $ \dot{v}_j^{(m)} $. This leads to the fact that an ultrafast charge transfer between neighboring donor and acceptor molecules continues to oscillate without dephasing. Although it appears as a very thick curve in Fig.~\ref{fig:shift_evolution}(a), an ultrafast and large-amplitude oscillation is apparent if magnified on the time axis. In experiments performed at finite temperatures, such electronic motion is rapidly dephased. Indeed, if random numbers are introduced according to the Boltzmann distribution at a finite temperature of 0.01 eV in $ u_j $, $ v_j^{(m)} $, $ \dot{u}_j $, and $ \dot{v}_j^{(m)} $ of the initial state, such an ultrafast and large-amplitude oscillation disappears but the slow oscillations caused by $ u_j $ and $ v_j^{(m)} $ survive. The resultant evolution of $ \rho(t) $ becomes closer to that observed in ref.~\citen{uemura_prl10}. These fluctuations increase the ionicity on both sides of the phase boundary, which also becomes close to the experimentally observed behavior. As a consequence of the increased ionicity, the phase boundary is shifted to a larger $ \Delta $. We show an example of transient ionicity $ \rho(t) $ in Fig.~\ref{fig:random_shift_evol} in such a case of dephased charge transfer in a neutral state near the shifted boundary. 
The ultrafast and large-amplitude oscillation [the very thick curve in Fig.~\ref{fig:shift_evolution}(a)] is indeed suppressed here. 

The above facts indicate the presence of strings of lattice dimerization before the photoexcitation of the neutral phase in the experiment performed at 90 K above the neutral-ionic transition temperature, i.e., 81 K. \cite{uemura_prl10} They are also consistent with the X-ray diffuse scattering showing the characteristic length of 16 molecules for thermally induced ionic strings at 105 K. \cite{guerin_prl10} At 90 K, the characteristic length would be larger and already beyond the size for which we can treat the exact many-electron wave function. Such one-dimensional precursors locally break the inversion symmetry and make the quick growth and the $-$cosine-type oscillation of the dimerization possible. Considering the fact that the photoinduced neutral-to-ionic dynamics decays much faster than the photoinduced ionic-to-neutral dynamics, \cite{okamoto_prb04} the growth of such one-dimensional ionic domains is saturated before they form a three-dimensional ionic domain that is sufficiently metastable for longevity. 

The picture of the photoinduced neutral-to-ionic transition obtained here is different from the previous one, \cite{iwano_prl06} which suggested ultrafast growth of an ionic domain without dimerization, based on the static and regular neutral initial state without couplings to lattice phonons or to molecular vibrations. In reality, the neutral phase in equilibrium above the phase transition temperature contains strings of lattice dimerization, already before photoexcitation. Photoexcitation rapidly grows and oscillates the dimerization [Fig.~\ref{fig:shift_evolution}(b)], which enhances the charge transfer between neighboring molecules [Fig.~\ref{fig:shift_evolution}(a) compared with Fig.~\ref{fig:evolution}(a)]. In other words, precursors are transformed into ionic domains after photoexcitation. 

\begin{figure}
\includegraphics[height=9.6cm]{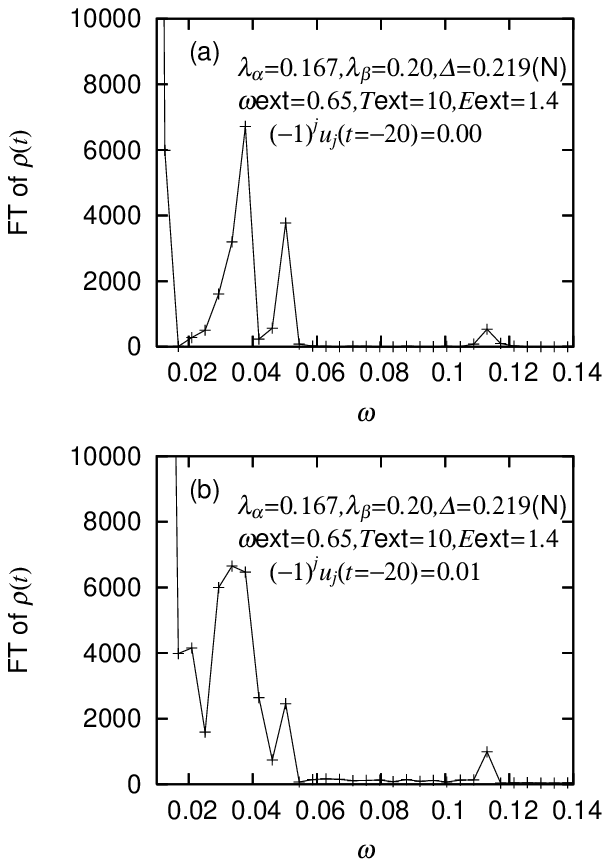}
\caption{
Fourier transform of ionicity after photoexcitation, (a) without and (b) with initial dimerization $ (-1)^j u_j (t=-20)=0.01 $ set artificially. 
\label{fig:fourier}}
\end{figure}

For clarity, the Fourier transform of $ \rho(t) $ during $ 50 < t < 1550 $ [longer than the period shown in Figs.~\ref{fig:evolution}(a) and \ref{fig:shift_evolution}(a)], defined by $ \mid \int_{50}^{1550} e^{i\omega t} \rho(t) dt \mid $, is plotted in Figs.~\ref{fig:fourier}(a) and \ref{fig:fourier}(b). 
The presence of a low-frequency component at $ \omega < $0.02 in Fig.~\ref{fig:fourier}(a) is due to the slowly growing dimerization contributing to $ \rho(t) $ after the period shown in Fig.~\ref{fig:evolution}(a). When the dimerization is introduced in the initial state, the dimerization oscillates from the beginning [Fig.~\ref{fig:shift_evolution}(a)] and contributes more to the low-frequency component in Fig.~\ref{fig:fourier}(b). With the exception of such a low-frequency component, Figs.~\ref{fig:fourier}(a) and \ref{fig:fourier}(b) are quite similar. All of the molecular vibrations evidently contribute to $ \rho(t) $: a peak at $ \omega $=0.038 due to $ v_\mathrm{A1}(t) $ with $ \omega_{\beta\mathrm{A1}} $=0.040, a peak at $ \omega $=0.050 due to $ v_\mathrm{D}(t) $ with $ \omega_{\beta\mathrm{D}} $=0.055, and a peak at $ \omega $=0.11 due to $ v_\mathrm{A2}(t) $ with $ \omega_{\beta\mathrm{A2}} $=0.12. All of them oscillate in the same phase as that in Figs.~\ref{fig:evolution}(b)-\ref{fig:evolution}(d). 

\section{Conclusions}

Neutral-ionic transition and photoinduced dynamics have intensively been studied for the mixed-stack charge-transfer complex TTF-CA. The importance of the coupling to lattice phonons has been recognized from the dimerization and the consequent ferroelectric order in the ionic phase. The roles of the couplings to molecular vibrations have not been paid much attention to, except for a few theoretical studies. \cite{painelli_prb88,soos_prb07} They are found to be important for the ionic phase to be a Mott insulator with large ionicity. The couplings to molecular vibrations stabilize the neutral phase, making the ionicity in the ionic (neutral) phase near the boundary large (small). The couplings to lattice phonons (molecular vibrations) reduce (enhance) the discontinuities in physical quantities. Both couplings are necessary for the reproduction of the experimentally observed ionicity, optical conductivity, and photoinduced charge dynamics. 

The photoinduced ionicity dynamics also shows a large contribution from molecular vibrations. The comparison with the experimentally observed, photoinduced charge dynamics indicates the presence of strings of lattice dimerization as local and precursory symmetry breaking in the neutral phase above the transition temperature. Only when a small but finite dimerization is introduced in the initial state can we reproduce its $-$cosine behavior and its large contribution to the photoinduced ionicity. 

\section*{Acknowledgment}

This work was supported by Grants-in-Aid for Scientific Research (C) (Grant Nos. 19540381 and 23540426), Scientific Research (B) (Grant No. 20340101) and Scientific Research (A) (Grant No. 23244062), and by ``Grand Challenges in Next-Generation Integrated Nanoscience" from the Ministry of Education, Culture, Sports, Science and Technology of Japan, and the NINS program for cross-disciplinary study (NIFS10KEIN0160).

\bibliography{holNIa_2}

\end{document}